\documentclass{elsart}
\usepackage{epsfig}
\usepackage{rotate}
\usepackage{amssymb}
\usepackage{amsmath}

\newcommand{\ops}{\ensuremath{\mathrm{o\text{-}Ps}}}
\newcommand{\pps}{\ensuremath{\mathrm{p\text{-}Ps}}}

\newcommand{\invdecay}{\ensuremath{o-Ps \to \mathrm{invisible}}}
\newcommand{\binvdecay}{\ensuremath{Br(\invdecay)}}

\begin{document}
\begin{frontmatter}
\title{\boldmath An Improved Limit on Invisible Decays of Positronium}

\vspace{.5cm}
\author[Zuerich]             {A.~Badertscher}
\author[Zuerich]             {P.~Crivelli}
\author[Zuerich]             {W.~Fetscher}
\author[Zuerich]             {U.~Gendotti}
\author[INR]                 {S.N.~Gninenko}
\author[INR]                 {V.~Postoev}
\author[Zuerich]             {A.~Rubbia}
\author[INR]                 {V.~Samoylenko}
\author[LAPP]                {D.~Sillou}

\address[Zuerich]        {ETH Z\"urich, Z\"urich, Switzerland}
\address[INR]            {Inst. Nucl. Research, INR Moscow, Russia}
\address[LAPP]           {Laboratoire Physique des Particules Annecy, LAPP, France}

%\clearpage
\begin{abstract}
The results of a new search for positronium decays into invisible final states are reported.  Convincing detection of this decay mode would be a strong evidence for new physics beyond the Standard Model (SM): 
for example the existence of extra--dimensions, of milli-charged particles, 
of new light gauge bosons or of mirror particles. Mirror matter could be a relevant dark matter candidate.
 
In this paper the setup and the results of a new experiment are presented. In a collected sample of about $(6.31\pm0.28) \times 10^6$ orthopositronium decays, no evidence for invisible decays in an energy window [0,80] keV was found and an upper limit on the branching ratio of orthopositronium \invdecay\ could be set:
\begin{equation}\nonumber
\binvdecay<4.2\times 10^{-7}~(90\%~\textrm{C.L.})
\end{equation}

 Our results provide a limit on the photon 
mirror-photon mixing strength $\epsilon \leq 1.55\times 10^{-7}~(90\%~\textrm{C.L.})$ and rule out particles lighter than the electron mass with a fraction  
$Q_x \leq 3.4 \times 10^{-5}$ of the electron charge. 
Furthermore, upper limits on the branching ratios  for the decay of parapositronium $Br(p-Ps\to invisible)\leq 4.3 \times 10^{-7}~(90\%~\textrm{C.L.})$ and the direct annihilation 
$Br(e^+e^-\to invisible)\leq 2.1 \times 10^{-8}~(90\%~\textrm{C.L.})$ could be set.

\end{abstract}
\end{frontmatter}

\section{Introduction}\label{sec:intro}
 
Although direct manifestations of new physics are searched for at the high energy frontier, new phenomena can also be looked for at low energies via precision measurements. The new effects might be observed in rare decays of the positronium (see e.g. \cite{wsproceedings}).

Positronium (Ps), the positron-electron bound state, 
 is the lightest known atom, which at the 
current level of experimental and theoretical precision is bound and self-annihilates 
through the electromagnetic interaction \cite{adkins}. 
This feature has made positronium an ideal system for  
testing the accuracy of Quantum Electrodynamics (QED) calculations 
for bound states, in particular for the triplet ($1^3S_1$)
state of $Ps$ (orthopositronium, \ops ) \cite{karshenboim2004}. 
Due to the odd-parity under
C-transformation  \ops\ decays
predominantly into three photons with a lifetime in vacuum of $\tau_{\ops }=142.05$ ns \cite{adkins}-\cite{powder2003}. 
 The singlet ($1^1S_0$) state (parapositronium, \pps ) decays predominantly into two photons with a lifetime in vacuum of $\tau_{\pps }=125$ ps \cite{czarnecki,ppsdecayrate}. 
 The longer lifetime of \ops\ (due to the phase-space and additional
$\alpha$ suppression factors) gives an enhancement factor $\simeq 10^3$
in the sensitivity to an admixture of potential 
new interactions not accommodated in the Standard Model (SM) \cite{matveev}.

This paper focuses on a new search for  \invdecay\ decays. By invisible we mean photonless decays, i.e. decays which are not accompanied by energy deposition in a hermetic calorimeter. 
In the SM the decay into a
 neutrino-antineutrino pair has a branching ratio of $6.6\times 10^{-18}$ \cite{karshenboim1999}. Evidence for invisible decays in the region $\simeq 10^{-7}$  would, therefore, unambiguously signal the presence of new physics. New models that are relevant to the $o-Ps\to invisible$ decay mode 
predict the existence either of i) extra-dimensions \cite{tinyakov,extradim2}, or ii) fractionally charged particles \cite{holdom,david}, or iii) a new light vector gauge boson \cite{extradim}, or
 iv) mirror particles, which could be candidates for dark matter \cite{ly}-\cite{okun}.

The first experiment to search for invisible decay channels of o-Ps was performed by Atoyan et al.~\cite{atoyan}. Their result  on \binvdecay\ $<5.3\times 10^{-4}$ ($90\%$ C.L.) excluded this channel as a possible explanation of the \ops\ lifetime anomaly (for a recent review see e.g. \cite{dsi}). This search was repeated by  Mitsui et al. who found a branching ratio \binvdecay\ $< 2.8 \times 10^{-6}$ ($90\%$ C.L.) \cite{mits}. Furthermore, they could place a  limit on the existence of milli-charged particles and on the photon mirror-photon mixing. This result was corrected in \cite{gninenko} by taking into account the suppression factor for the mixing due to the presence of matter.

The rest of the paper is organized as follows. The details of the experimental setup are reported in Section \ref{sec:experiment}. The expected background is presented in Section \ref{bkg}. In Sections \ref{sec:datareduction} and \ref{sec:results} the data analysis and the results are described. The interpretation and the conclusion are reported in Sections \ref{sec:interpretation} and \ref{sec:conclusion}.      

\section{Experimental technique and setup}\label{sec:experiment}

The experimental signature of \invdecay\ decay is the apparent disappearence of the energy  $2m_e$ expected in ordinary decays in a hermetic calorimeter surrounding the \ops\ formation target. The readout trigger for the calorimeter is performed by tagging the stopping of a positron in the target with high efficiency.

For the design of the experimental setup aiming at a sensitivity \binvdecay\ $\simeq 10^{-8}$, the following criteria were considered:

\renewcommand{\labelenumi}{(\alph{enumi})}
\begin{enumerate}
\item  The probability not to detect all direct  $e^+e^-$ annihilation photons was suppressed to $\le 10^{-9}$ using a thick hermetic crystal calorimeter, minimizing the dead material and with a good energy resolution. The probability to loose all photons in 3$\gamma$ decays is consequently even smaller.
\item The region around the target was designed with as little dead material as possible in order to reduce photon losses.
\item By the appropriate choice of a porous target material and its dimensions a high fraction of \ops\ was produced resulting in a suppression of the background from the the direct  $e^+e^-$ annihilation and \pps\ decays and in high  statistics.
\item The trigger rate and the DAQ speed were maximized for statistics.
\item An efficient positron tagging system was designed to provide a clean trigger for positronium formation. A method was developed to suppress the background from the electrons emitted in the EC process (shake-off electrons). 
\item An efficient identification of the 1.27~MeV photon emitted by the $^{22}$Na radioactive source was achieved with a method to veto the charged particles entering the trigger counter, thus, the backgrounds related to them could be reduced.  
\end{enumerate}

\begin{figure}[h!]
\hspace{.0cm}\includegraphics[width=.6\textwidth]{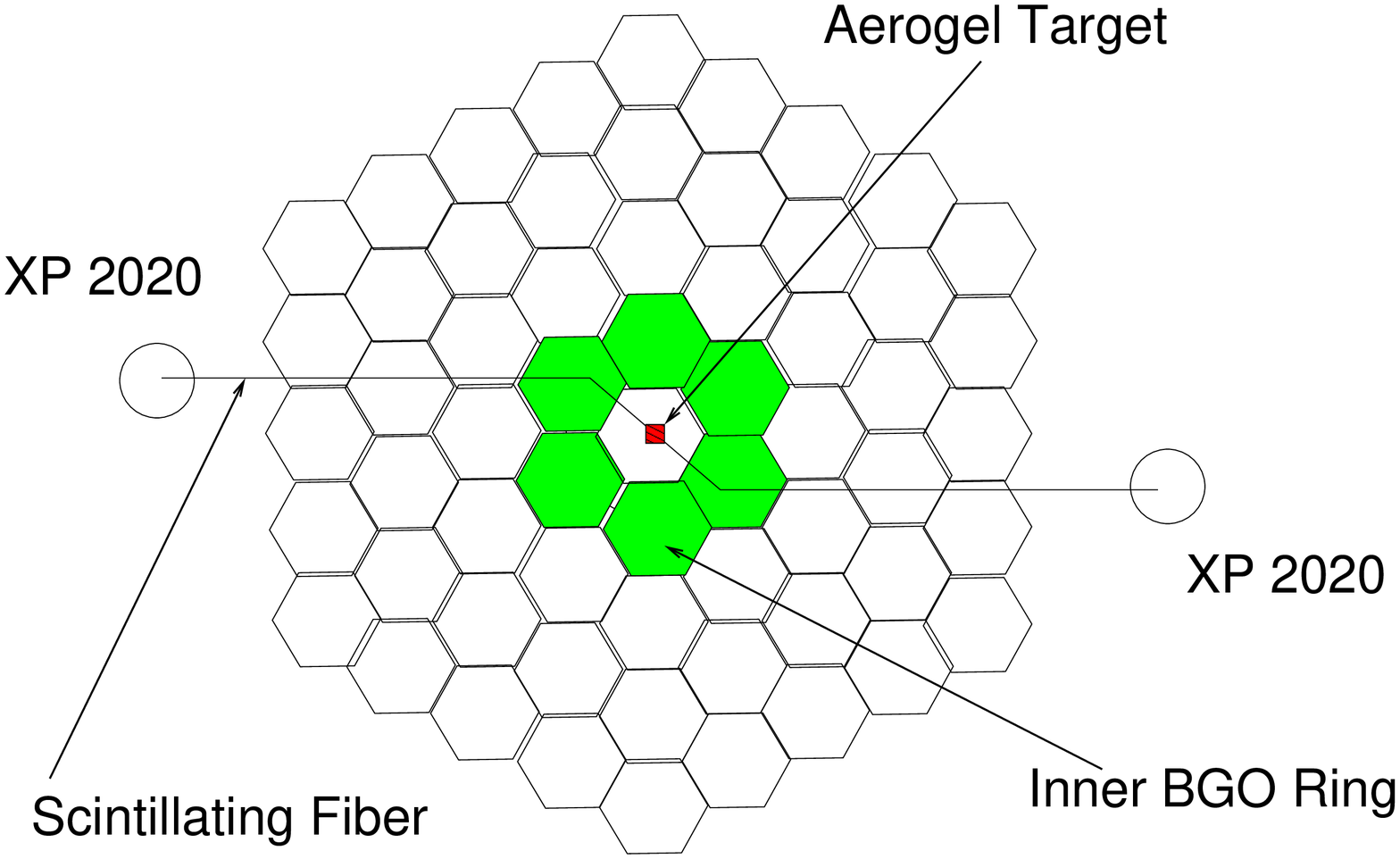}
\hspace{1.cm}\includegraphics[width=.4\textwidth]{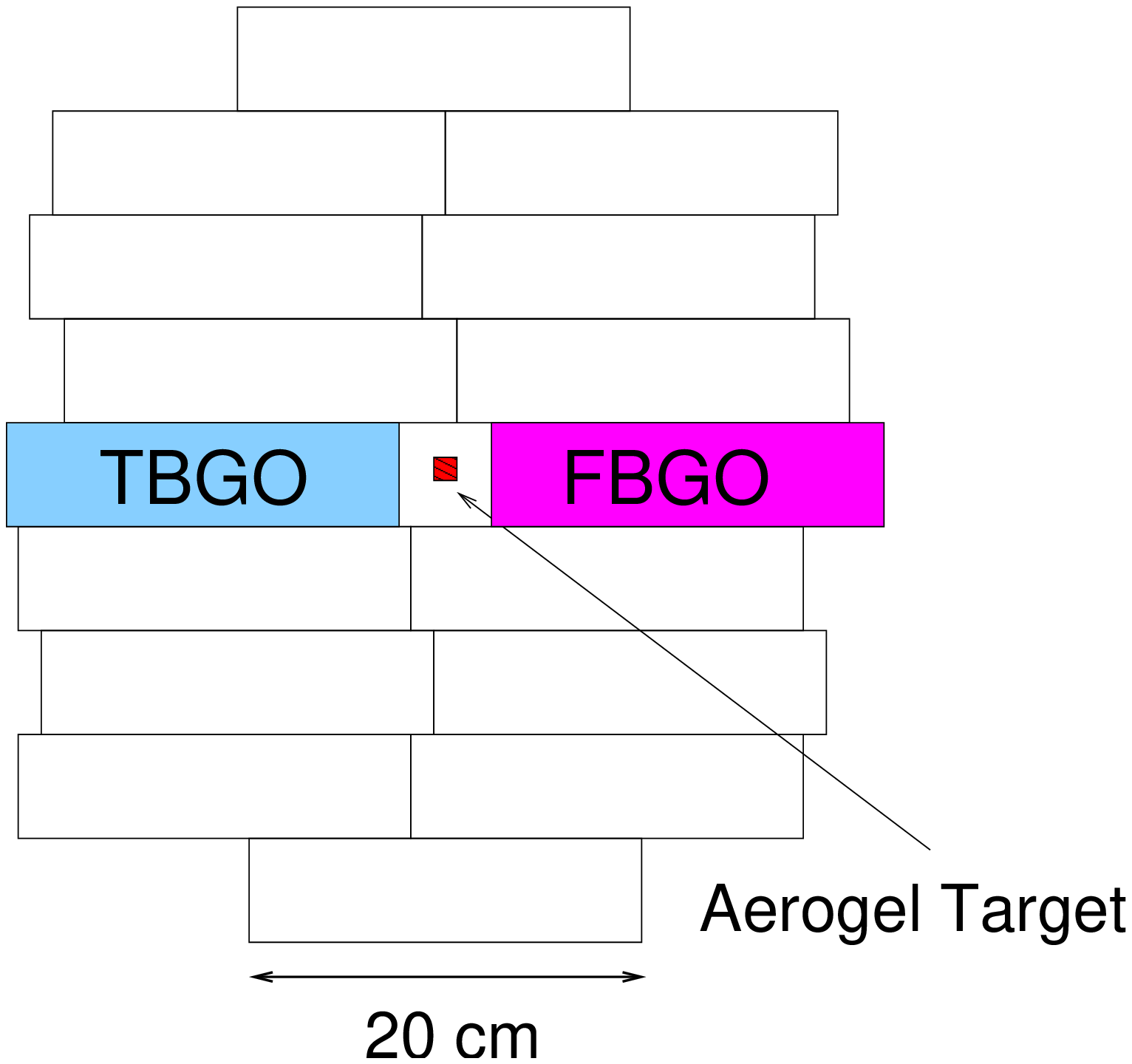}
\vspace*{-.cm}
\caption{\em Schematic illustration of the experimental setup: a) front view, b) top view.}
\label{detector}
\end{figure}

The schematic illustration of the detector setup is shown in Figure \ref{detector} (the detailed description of the experimental technique and setup can be found in Refs. \cite{pol,thesis}). 

\begin{figure}[h!]
\hspace{.0cm}\includegraphics[width=\textwidth]{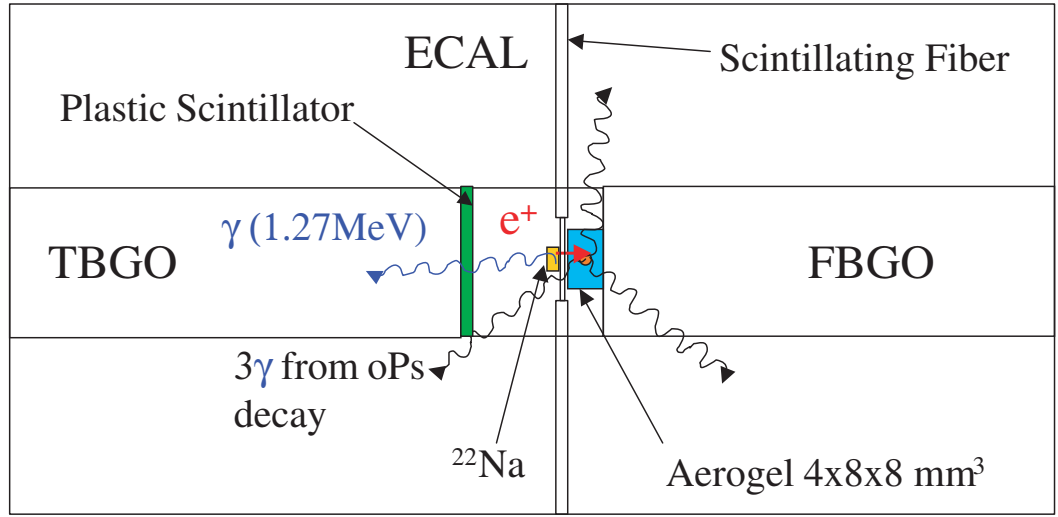}
\vspace*{.0cm}
\caption{\em Schematic illustration of the positron tagging system and the \ops\ formation target of the setup.}
\label{opsregion}
\end{figure}

 Positrons are produced from a $^{22}$Na source 
with an activity of $\simeq 30$ kBq. The $^{22}$Na has a half life of 2.6 years and has a $Q$-value for the nuclear transition to  $^{21}$Ne of $Q=2.842$~MeV. This is the maximum energy available for the particles involved in one of the three possible decay modes of $^{22}$Na: 
\begin{enumerate}
\item Decay mode A (Br $\simeq 90.6\%$): the $\beta ^+$ decay with end-point energy 0.546~MeV. The positron is always followed by the prompt emission of a 1.27~MeV photon ($\tau \simeq$ 3.7 ps) from the $^{21}$Ne$^*$ de-excitation to the ground state. 

\item Decay mode B (Br $\simeq 9.44\%$): the Electron Capture process (EC), where an orbital electron is captured by the nucleus, and only a 1.27~MeV photon and a neutrino are emitted from the source. In some rare cases (with a probability  $\simeq 6 \times 10^{-3}$), an orbital electron  is ejected, due to the sudden change in the nucleus charge (shake-off) \cite{shakeoff}.  

\item Decay mode C (Br $\simeq 0.056\%$): there is no photon emission because the transition goes directly to the ground state. The end point energy of the positron is 1.83~MeV. 

\end{enumerate}
Therefore, in most \ops\ (\pps) decays we expect 3(2) photons with a summed energy equal to $2m_e$ and one photon with an energy of 1.27~MeV (see Figure~\ref{opsregion}). 

Photons  from the direct $e^+e^-$ annihilation in flight or from the positronium decays were detected in a hermetic, segmented BGO calorimeter (the ECAL). Two endcap counters called TBGO and FBGO (see Figures~\ref{detector}-\ref{opsregion}) surrounded the target on each side. At the analysis level the 1.27~MeV photon (``the triggering photon'') was required to be identified in the TBGO counter. The calorimeter was instrumented with charge and time readout.

\begin{figure}[h!]
\hspace{.0cm}\includegraphics[width=\textwidth]{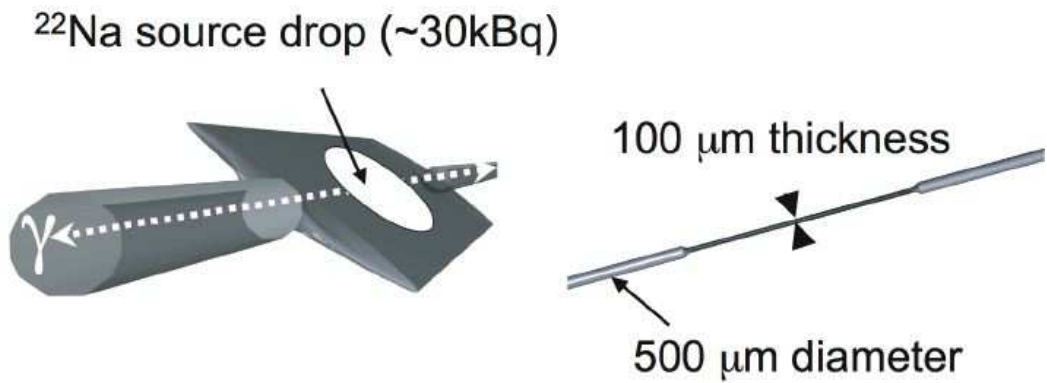}
\vspace*{.0cm}
\caption{\em Schematic view of the scintillating fiber with the $^{22}$Na source on the squeezed part in the center.}
\label{fiber}
\end{figure}

The activity of the source was chosen to maximize the trigger rate versus the inefficiency of signal detection (mostly due to pileup events). The source was prepared  by evaporating drops of a $^{22}$Na solution directly on a 100 $\mu$m thick and 2x8~mm$^2$ wide plastic scintillator (see Figure~\ref{fiber}) fabricated by squeezing a 500 $\mu$m diameter scintillating fiber (Bicron BF-12). In this way, no dead material was introduced for a source holder. The S-shape of the fiber (see Figure~\ref{detector}) was selected to avoid background from back-to-back 511~keV annihilation photons. 
The scintillating fiber was read out at both ends by two photomultipliers
(Philips XP2020) located outside the detector (see Figure~\ref{detector}). 
The coincidence of the two PMT signals was used to
tag the passage of a positron through the fiber and acted as a start signal for the data readout system. The use of two PMTs in coincidence, instead of a single one \cite{mits}, lowered the ratio between fake and real positron triggers to $<1.9\times 10^{-10}$.

\begin{figure}[h!]
\hspace{.0cm}\includegraphics[width=\textwidth]{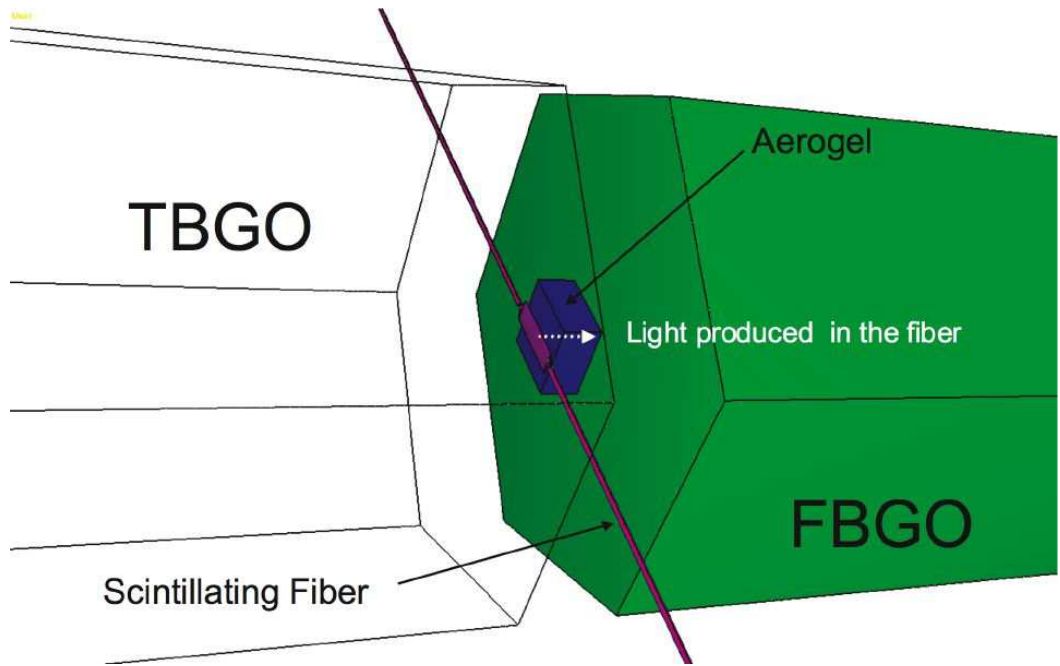}
\vspace*{.0cm}
\caption{\em Schematic illustration of the method to readout the energy deposition in the fiber.}
\label{fbgofiber}
\end{figure}

Opposite to the source, a 4x8x8~mm$^3$  SiO$_2$ aerogel piece (type SP30, purchased from Matsushita Electric Works) was placed in contact with the squeezed fiber (see Figure~\ref{opsregion} and \ref{fbgofiber}).
Positrons stopping in the aerogel
target may form positronium (the formation probability is $45\%$ \cite{kalimoto}) which can migrate into the aerogel pores. The collisions with the walls of the pores did not appreciably quench the \ops: when the aerogel was flushed with nitrogen, a fit to the distribution of the time difference between the start from the fiber and the stop from the calorimeter yielded $\tau_{\ops}=129.1\pm1.8$ ns, which is very close to the lifetime in vacuum $\tau_{\ops}\simeq 142.05\pm0.02$ ns \cite{PDG}.

The ECAL was composed of 100 BGO crystals that 
surrounded the target region providing a nearly isotropic sphere of radius 200-220~mm (see Figure~\ref{detector}). 
Each crystal had a hexagonal cross-section of 
61~mm diameter and a length of 200~mm. The crystals of the inner most ring were wrapped in a 2 $\mu$m thick aluminized Mylar foil to minimize the photon energy absorption. The other crystals were wrapped with Teflon foils of 750 $\mu$m thickness. The crystals in the barrel and the FBGO were readout with ETL 9954 photomultiplier tubes.

The energy deposited in the fiber is an important parameter in order to reject the background from the electrons emitted in the EC process (shake-off electrons, see Section \ref{bkg}). 
The mean number of photoelectrons detected in each XP2020 for a  positron crossing the fiber was measured to be about 1.2, thus, a cut on the energy deposited in the fiber using these signals would not be meaningful. For this reason the FBGO was also used to measure the energy deposited in the fiber by the positron:  the light emitted by the fiber could  traverse the transparent aerogel and enter the FBGO through an aperture in the wrapping on its front face. This light was then guided by the FBGO to the PMT attached on the back face of FBGO (as illustrated in Figure~\ref{fbgofiber}). This method provided a mean number of photoelectrons equal to 13$\pm1$ for a positron traversing the fiber \cite{thesis}.

 The TBGO endcap was used in the off-line analysis to identify the 1.27~MeV photon. The energy resolution of the TBGO was an essential parameter to reduce backgrounds related to the misidentification of the 1.27~MeV photon (see also Section \ref{bkg}). To provide a better energy resolution, this crystal was coupled to an ETL 9964 PMT with a more uniform light collection and a larger quantum efficiency than the ETL 9954. For the same reason this crystal was of a better quality than the others and efforts were dedicated to maximize the light collection by keeping the amount of dead material introduced by the crystal wrapping as small as possible. The best results were achieved with the crystal wrapped in the 3M radiant mirror (64 $\mu$m thickness): the resolution at 662~keV was measured to be about $15\%$ (FWHM). To select the triggering photon, an energy window $[1275\pm 67]$~keV was used in the analysis. In addition, to veto charged particles (positrons and electrons) entering the TBGO, a 1~mm thick plastic scintillator (Bicron BC-400) was optically coupled to the TBGO front face (for more details see \cite{sauter}), i.e. the same PMT was used to detect the light signals from the plastic scintillator and the TBGO crystal (see Section \ref{bkg} for a discussion of backgrounds associated to charged particles entering the TBGO). The signals from the plastic scintillator and the BGO could be distinguished because of the different decay time ($\tau$ = 2.7~ns for the plastic scintillator and  $\tau \simeq$ 300~ns for the BGO). For this purpose, split signals from the PMT were fed into two ADCs with different integration gates (called short gate and long gate).

A VME system interfaced to a PC was used for data
acquisition (the DAQ rate was about 1800 events/s).  For every trigger, five CAEN 32 channels QDC v792 modules recorded the charge of the crystal signals while a CAEN TDC v775 recorded the time information. A trigger gate length of 2.9 $\mu$s was chosen to keep the probability of \ops\ to decay after this time to $\lesssim 10^{-9}$.

A temperature stabilized light-tight box containing the calorimeter was built in order to keep the temperature of the BGO crystals in the range of $\pm 0.5~^0$C. Water, whose temperature was controlled by two thermostats, circulated through copper tubes welded on two copper plates inside the box. The experimental hall was air conditioned to keep temperature variations within $\pm 1^0$C. 
The high-voltage dividers of all PMTs  were placed outside the box  in order to avoid energy dissipation close to the crystals. 
The BGO crystals were equipped with LEDs that could be pulsed periodically to monitor the response. Additionally, the gains of the PMTs were also monitored to check their stability.

The detector was calibrated and monitored internally 
using the 511~keV annihilation photon  
and the 1.27~MeV photons emitted by the $^{22}$Na source.
Variations of the energy scale during the run period were within $\lesssim 1\%$ and corrected on the basis of an internal calibration procedure \cite{thesis}.
 
\section{Background estimation and dedicated engineering run}\label{bkg}
In order to reach the required sensitivity, the background must be reduced and controlled at the level of $10^{-8}$.
To understand the different background sources and to cross-check the simulation, we perfomed an engineering run with a simplified version of our detector. During two months of data taking, the stability of the detector and its components was investigated. The comparison between the background expected based on Monte Carlo (MC) simulations and the data of the engineering run is summarized in Table~\ref{Bkgtbl}.  

The ECAL thickness of 200~mm provides a probability of $<10^{-9}$ for two 511~keV photons to escape detection (see background 1 in Table~\ref{Bkgtbl}). For three photons decays, this probability is consequently even smaller. 

If one (or more) annihilation photon (e.g.  backscattered from the target) overlaps with the 1.27~MeV in the TBGO, it can fall in the trigger energy window $[1275\pm 67]$~keV because of the finite energy resolution. This introduces a background if the remaining annihilation photon gets absorbed in the dead material or escapes detection. The separation between the upper bound of the trigger energy window and the sum of a triggering and a 511 keV photon was, thanks to the good energy resolution of the TBGO, 7$\sigma$ of the 1786 keV peak (1275+511 keV). Thus, the level of this background is $<5\times10^{-9}$ (background 2 in Table~\ref{Bkgtbl}). 

 In order to suppress the other sources of background related to the misidentification of the 1.27 MeV photon (backgrounds 3, 4 and 5 listed in Table~\ref{Bkgtbl}), one had to veto charged particles entering the TBGO.
Two processes are responsible for generating such triggers.
One is associated with decay mode B (EC) when the 1.27 MeV photon interacts in the fiber, faking a positron signal. If the scattered photon and the Compton electron reach the TBGO, the sum of the energy of the two particles can be misidentified as the triggering photon without any energy in the rest of the calorimeter.
Another background can occur in decay mode A if the 1.27 MeV photon is not detected: a trigger can be produced by a positron that multiple scatters (MS) in the fiber and deposits enough energy to trigger the experiment. If the positron reaches the TBGO with a kinetic energy of about 200-300 keV and the  two 511 keV annihilation photons are completely absorbed in the TBGO an energy close to  1.27 MeV  will be reconstructed. This will appear as an invisible decay since no energy is expected in the rest of the detector.
If the 1.27 MeV photon is not present (decay mode C) a trigger can similarly  be produced by the 1.83 MeV positron. To veto these backgrounds the charged particle veto of the TBGO described in the previous section was used such that the backgrounds 3, 4 and 5 had a probability of $<10^{-8}$ in the final setup. 

The EC photon may accidentally coincide with a trigger from the fiber generated either by the PMTs noise or by some other particles emitted from possible unstable isotopes formed during the target activation. The level of this background  could be reduced with the selection of two XP 2020 with very low noise ($<30$ counts/s) and the requirement of the coincidence between them. In addition, a radioactive source with a controlled high purity was chosen. From the data of the engineering run, this background  was estimated to be $<1.9\times 10^{-10}$ (background 6 in Table~\ref{Bkgtbl}).       

During the decay mode B (EC) the fiber signal can be generated by shake-off electrons (background 7 in Table~\ref{Bkgtbl}). Since the probability of an electron ejection steeply drops with its emission kinetic energy (more than 4 orders of magnitude in the first 100 keV) the cut on the energy deposited in the fiber by the triggering particles was used to suppress this background. 

The engineering run allowed to test these backgrounds as shown in Table~\ref{Bkgtbl}. The expected fraction of zero energy events is 10$\%$ smaller than what was measured. This difference can be explained by the contribution from the shake-off electrons which was not included in the simulation. Indeed, in the engineering run there was no information about the energy deposited in the fiber so that no cut on the energy of the particles passing through the fiber could be applied.  
   
The last column of the table lists the expectations for the final setup. The total background is estimated to be at the level $10^{-8}$.
The threshold on the energy deposited in the fiber was set considering the uncertainty due to the number of photoelectrons and the subtraction method to be sure that no electron below 100 keV could trigger the fiber; it was chosen, based on simulations, to be at 140 keV.

\begin{table}[!h]
    {\begin{tabular}{|c|c|c|c|c|} \hline
	\multicolumn{2}{|c|}{BACKGROUND} & \multicolumn{2}{c}{ENGINEERING RUN} \vline&  \multicolumn{1}{c}{FINAL SETUP}\vline\\
	\multicolumn{2}{|c|}{SOURCE}	&   expected & measured  & expected \\
	\hline
	\hline

	1)&Hermiticity\hphantom{00} & & & \\
	&Dead Material\hphantom{00} & $<10^{-9}$ &$<10^{-9}$ &  $<10^{-9}$ \\
	&Resolution\hphantom{00} & & &  \\
	\hline
	2)&Absorption in trigger & & & \\ 
	&Energy window\hphantom{00} & $1.3 \times 10^{-6}$ &$1.5 \times 10^{-6}$  & $<5\times 10^{-9}$\\
	\hline
	3)&MS positron & & & \\ 
	& with $\mathrm E_{max}$=546keV\hphantom{00} & $2.1 \times 10^{-6}$ & &$<10^{-8}$ \\
	\hline
	4)&MS positron & & & \\ 
	& with $\mathrm E_{max}$=1.83MeV\hphantom{00} & $1.4 \times 10^{-7}$ & &$<10^{-8}$ \\
	\hline
	5)&Compton EC photon\hphantom{0} & $1.3\times 10^{-6}$ & &$<10^{-8}$\\
	\hline
	6)&Accidental noise \hphantom{0} &  & & \\
	&and EC photon\hphantom{0} & $3.2\times 10^{-11}$ &$<1.9\times 10^{-10}$ &$1.9\times 10^{-10}$ \\
	\hline
	7)&Shake--off electrons & & & \\ 
	&in EC process\hphantom{0} & $10^{-6}$-$10^{-7}$  & &$10^{-8}$ \\
	\hline
	\hline
	\multicolumn{2}{|c|}{Total}& $4.8\times10^{-6}$  & $5.6\times10^{-6}$ & $10^{-8}$ \\
 	\hline
    \end{tabular}}
    \caption{\em Comparison between expected and measured background level for the different background sources in the engineering run and expected background level for the final setup (see text for details).}
  \label{Bkgtbl}
\end{table}

\section{Data analysis}\label{sec:datareduction}  

For the analysis we used a data 
sample  of  $1.39 \times 10^{10}$ recorded fiber triggers collected over a four months data taking period. 
For each event 
the following variables were used and cuts were applied to suppress background:  

\renewcommand{\labelenumi}{(\arabic{enumi})}

\begin{enumerate}

\item $ \Delta T_{\textrm{short}}$ is the time from the trigger start to the end marker of the dual timer unit that generates the short gate (measuring the light from the plastic scintillator coupled to the TBGO).

\item $T _{\textrm{long}}$ is the pedestal of one of the QDC channels integrated using the long gate as a start trigger.

\item $ \Delta T _{\textrm{XP}}$ is the time difference between the two XPs reading the fiber.
\item $ \Delta T _{\textrm{TBGO}}$ is the time difference between the XPs coincidence and the TBGO.
\item $ E_{\textrm{TBGOc}}$ is the energy deposited in the TBGO with a gate delayed by 15~ns to measure the energy deposited in the TBGO crystal without the contribution of the plastic scintillator. 
\item $ E_{\textrm{TBGO}}$ is the energy deposited in the TBGO with the full gate (long gate). 
\item $ E_{\textrm{FBGO}}$ is the energy deposition in the fiber measured with the FBGO. 

\end{enumerate}

\begin{figure}[!h]
\begin{center}
\includegraphics[width=.45\textwidth]{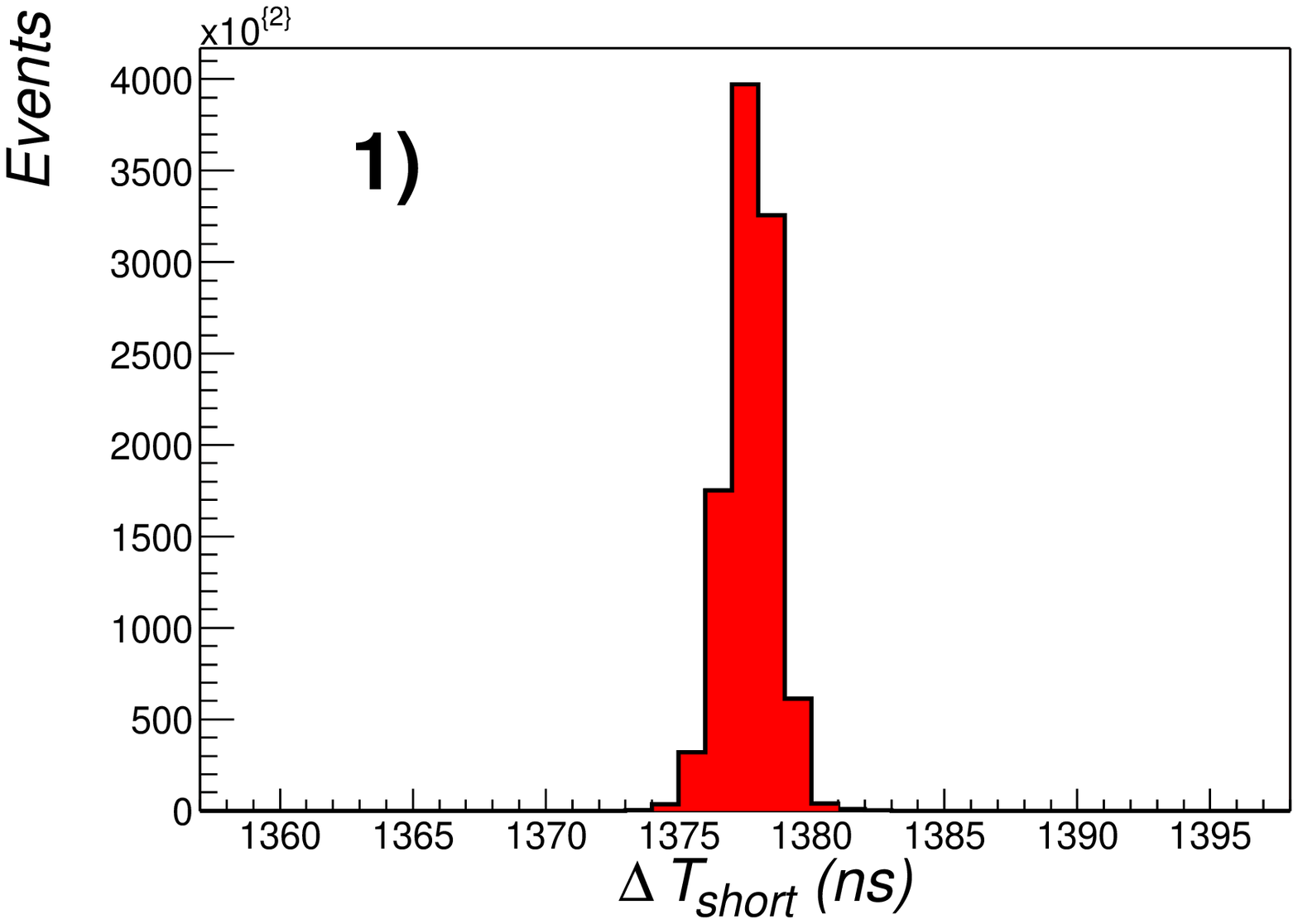}
\includegraphics[width=.45\textwidth]{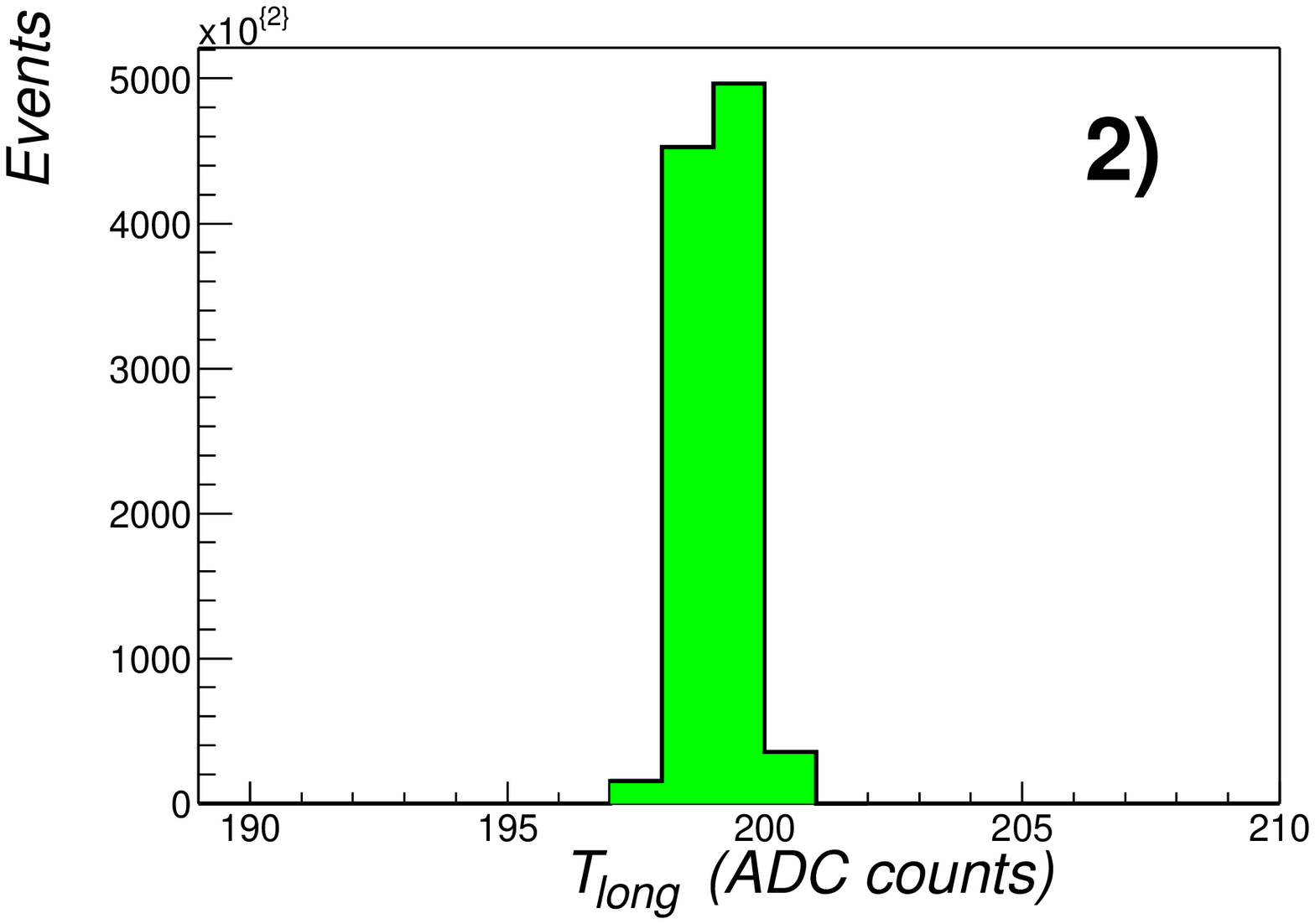}
\includegraphics[width=.45\textwidth]{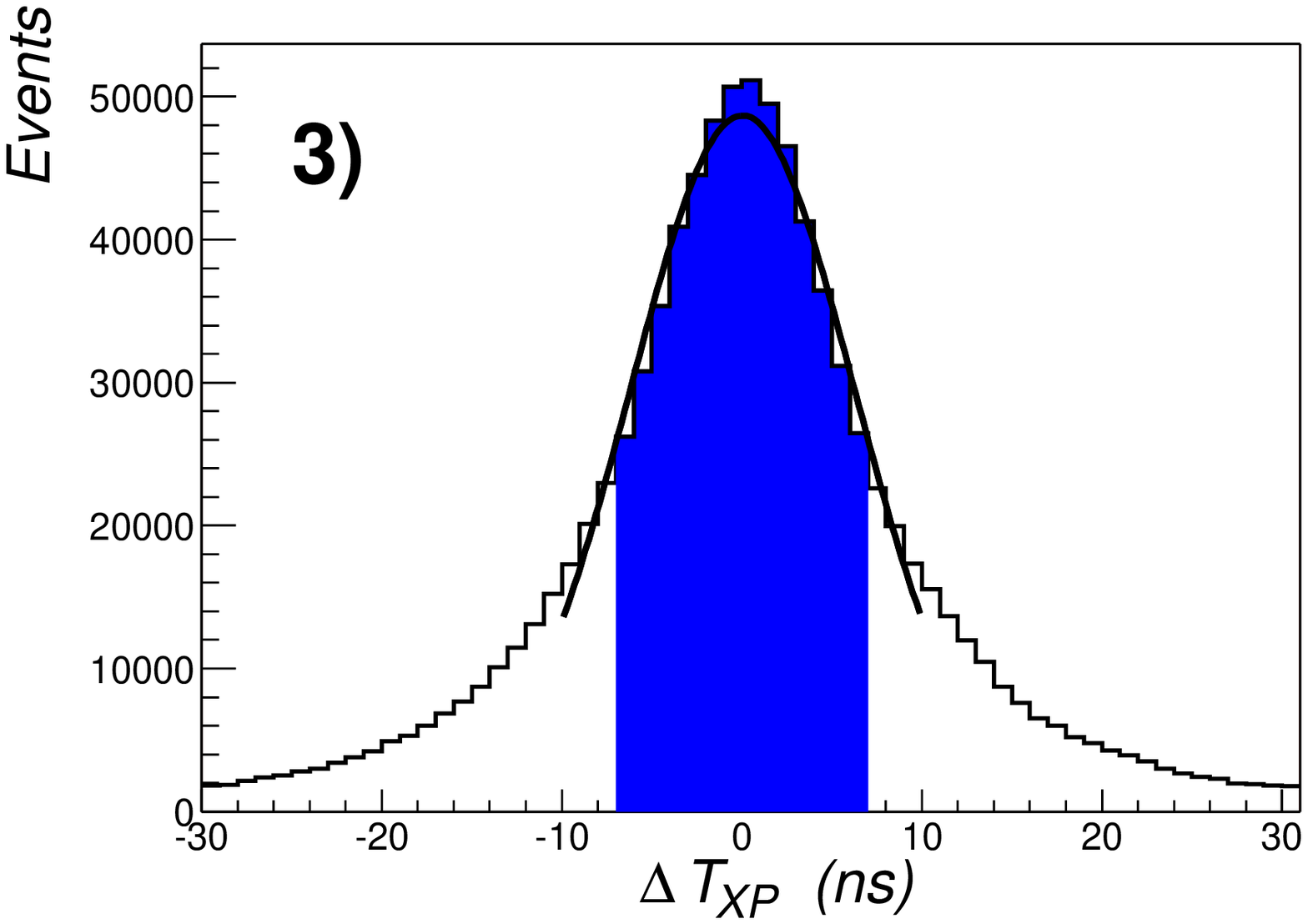}
\includegraphics[width=.45\textwidth]{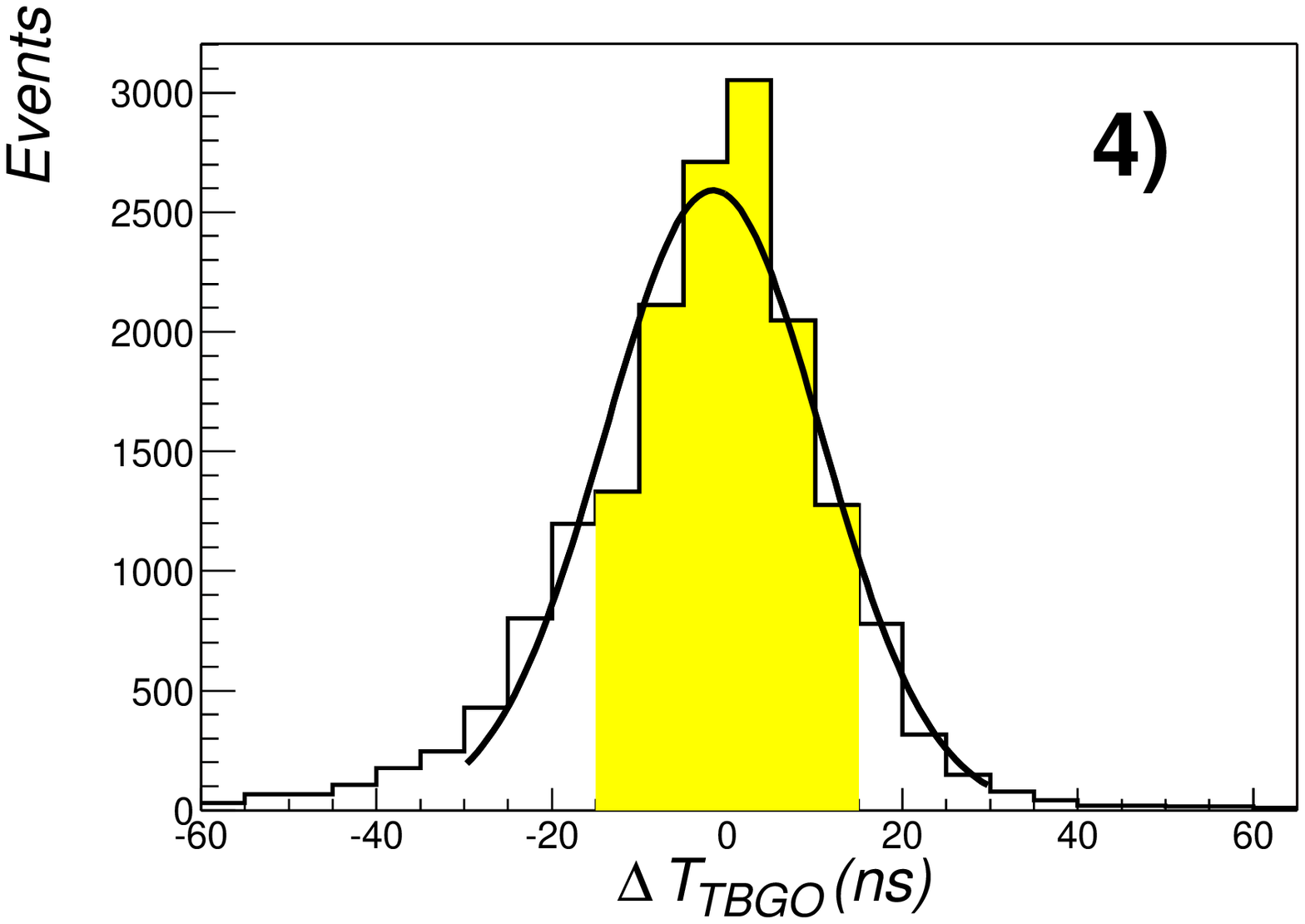}
\includegraphics[width=.45\textwidth]{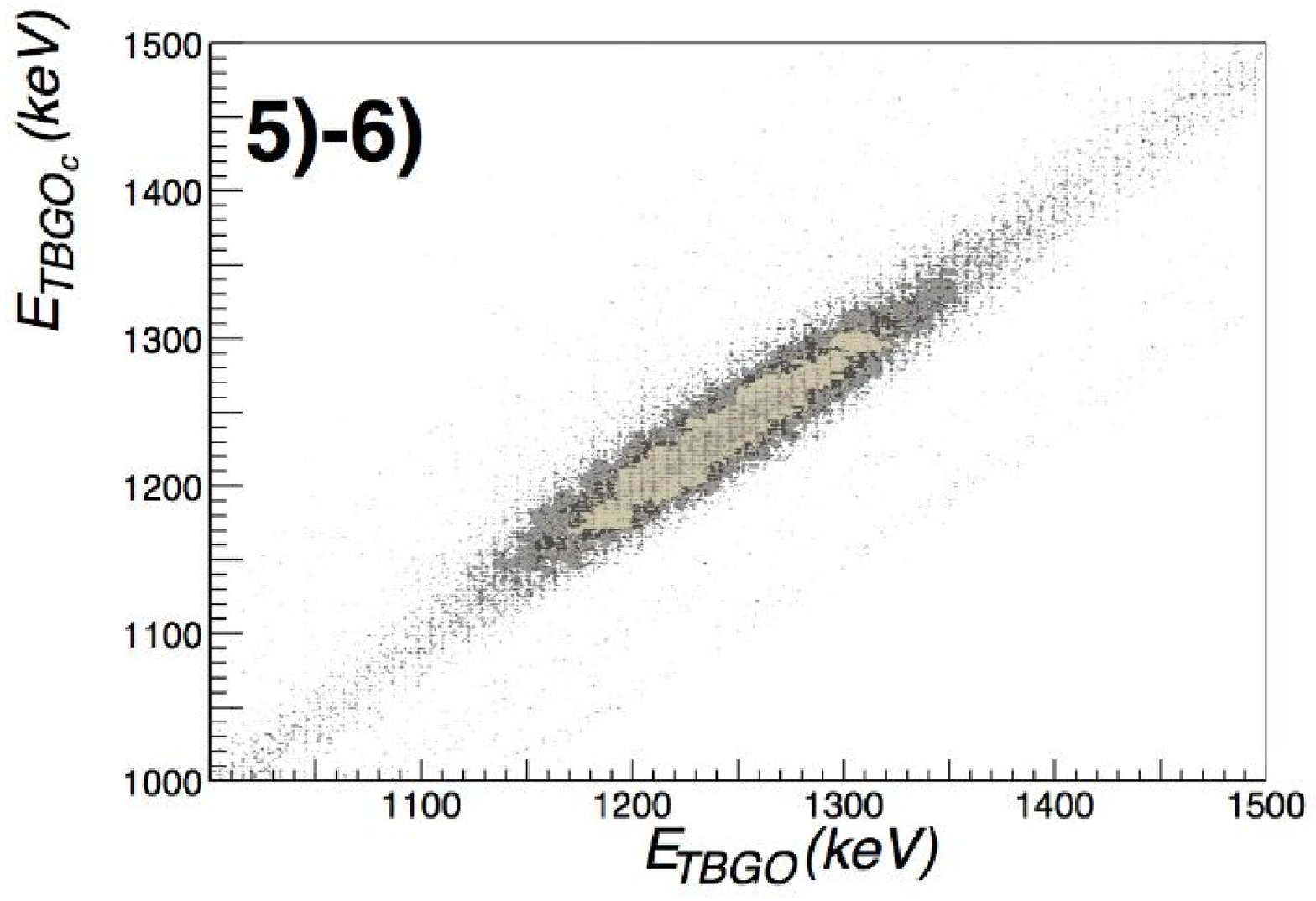}
\includegraphics[width=.45\textwidth]{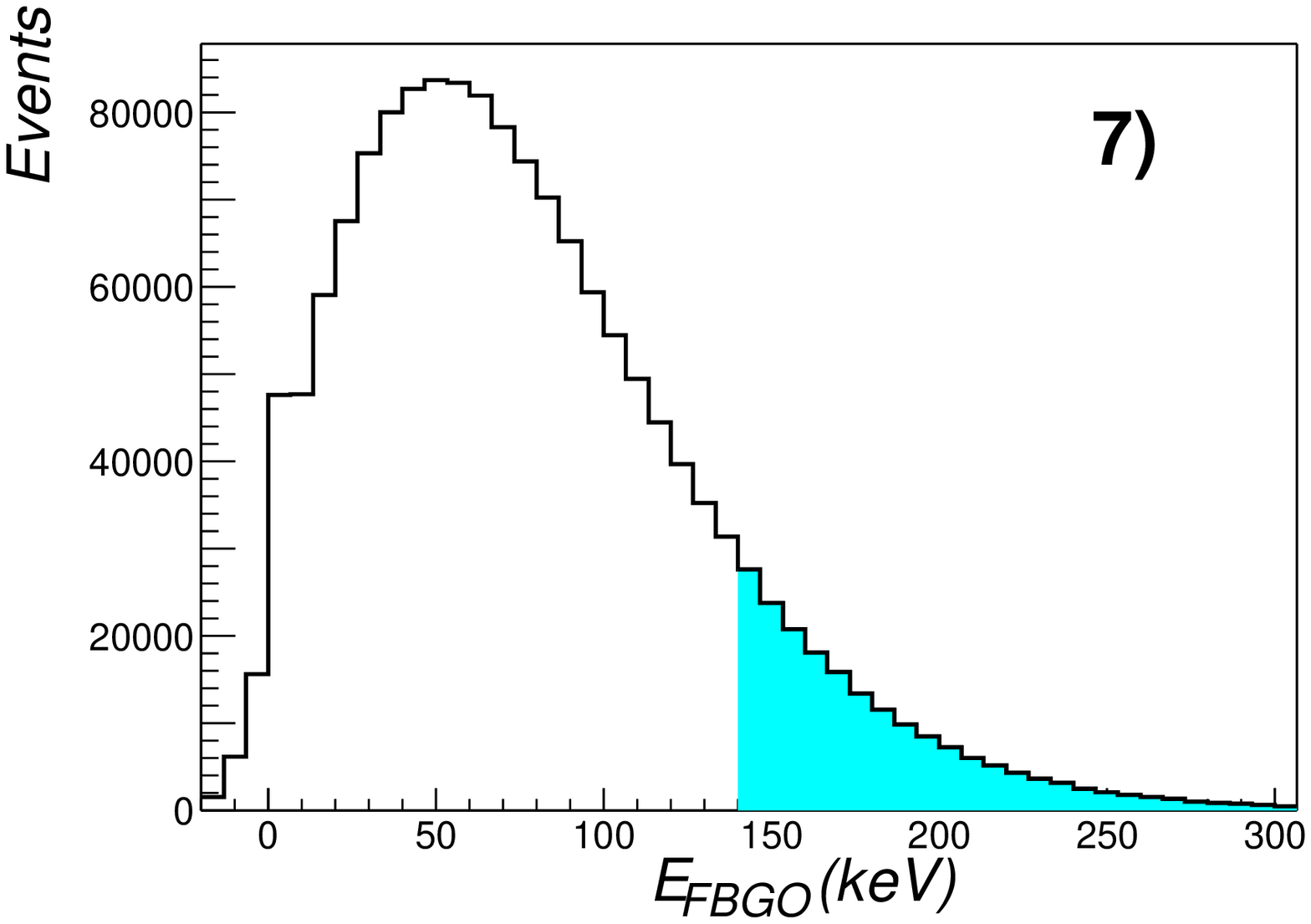}
\caption{\em   The cuts applied to the variables. The numbers on the plot correspond to the variable defined in the text. Only the colored regions contribute.}
\label{cuts}
\end{center}
\end{figure}

 The distributions of these variables for a reduced data sample of $10^{6}$ triggers are shown in Figure \ref{cuts}. The used selection cuts are listed in Table \ref{tab:cuts2005}. The cuts were selected and tuned with the help of a dedicated run with corresponding statistics of about $5\%$ of the data.       
These variables can be grouped in three categories, depending on their function: (a) The first two variables check the stability of the electronics and the duration of the gate widths. The selection has been tuned experimentally looking at the obtained spectra. (b) The variables 3) and 4) suppress accidental triggers faking positrons in the fiber. The cuts have been chosen to minimize the accidentals and maximize the signal statistics.
(c) The variables 5) to 7) are the cuts that reduce triggers from backgrounds that mimic the appearance of a positron in the formation cavity region.
 Furthermore, the upper limit of the energy window for the 1.27 MeV photon is very sensitive to the background from the ``absorption'' of one 511 keV $\gamma$ in the trigger energy window. Therefore, the long gate integration, which has a better resolution, is used to define this selection, while the short gate is used to reject the events with some energy deposited in the scintillator. The cut of $\pm 1\sigma$ was selected in order to enhance good triggers to the required level. 

All the selection cuts, except 7), have been defined in terms of the sigma of the signal determined with a Gaussian fit to the data sample with 10$^6$ events. Table~\ref{tab:cuts2005} summarizes the values used and the evolution of the total number of events passing the cuts.
 
The lower cut for the energy deposited in the fiber has been chosen to reduce the probability of shake-off electrons to trigger the fiber to a level $<10^{-8}$ as discussed in Section \ref{bkg}. 
The measured fraction of the \ops\ produced in the aerogel is reduced by $20\%$ applying the threshold 
for the energy cut in the fiber. This was expected, since the positrons that deposit the most energy are the ones stopping in the fiber.
 
\begin{table}[!h]
  \begin{center}
    {\begin{tabular}{|l|c|c|c|}
	\hline
	Variable & \multicolumn{2}{|c|}{Selection cut} & Fraction of events   \\
	name & $\#$ $\sigma$'s & value of 1$\sigma$         & remaining after cuts \\
	\hline
	1) $\Delta T_{short}$ & $\pm 4\sigma$ & 1.03 ns        & 99.3$\%$\\
	2) $ T_{long}$  & $\pm 4\sigma$ & 0.8 ADC counts & 98.9$\%$\\
	3) $\Delta T_{XP}$    & $\pm 1\sigma$ & 1.87 ns        & 75.5$\%$ \\
	4) $\Delta T_{TBGO}$  & $\pm 1\sigma$ & 3.71 ns        & 27.2$\%$\\
	5) $E_{TBGO}$  & $\pm 1\sigma$ & 74 keV         & 3.1$\%$\\
	6) $E_{TBGOc}$& $\pm 1\sigma$ & 67 keV         & 2.7$\%$\\
	7) $E_{FBGO}$  & \multicolumn{2}{|c|}{140 keV $<E_{FBGO}<$ 400 keV} & 1.1$\%$\\	\hline
    \end{tabular}}
    \caption{\em Definition of cuts and the remaining fraction of events after the cut is applied.}
    \label{tab:cuts2005}
  \end{center}
\end{table}

\section{Results}\label{sec:results}  

\begin{figure}[!h]
 \begin{center}
\includegraphics[width=.8\textwidth]{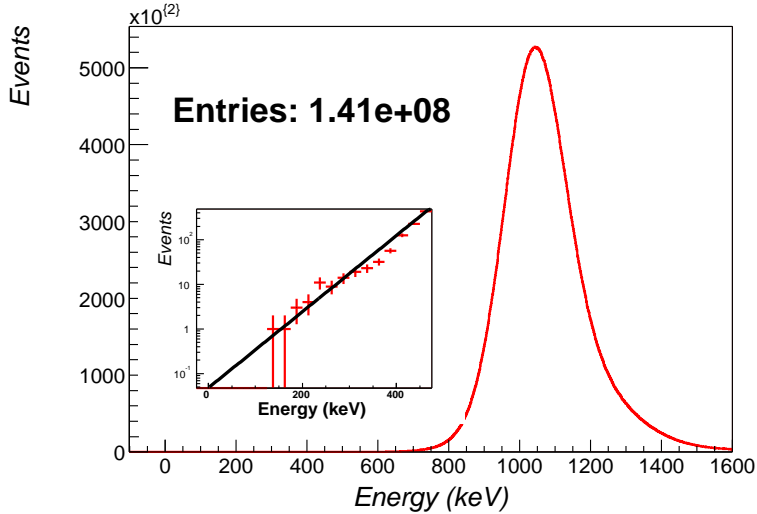}
\caption{\em  Spectrum of the sum of the total energy in the ECAL. The inset shows the magnified view of the low-energy region in logarithmic scale.}
\label{EcalSum}
 \end{center}
 \end{figure}

After imposing the above requirements a final sample of 1.41$\times 10^{8}$ events was obtained. 
For these events the energies of all the 100 BGO crystals, except the TBGO, were summed. Figure~\ref{EcalSum} shows the spectrum of the total energy ($E_{\textrm{tot}}$) deposited in the ECAL. The peak at 1022 keV corresponds to the positronium mass ($M_{Ps}\simeq 2m_e$). The inset shows that no event is observed in the zero energy region.

To define the upper energy cut on $E_{\textrm{tot}}$, below which an event is considered as photonless (invisible), a dedicated run of $10^7$ triggers was performed triggering the experiment only with the 1.27 MeV photon and no requirement of the fiber \cite{atoyan}. The zero peak contained about  10.4$\pm 1.2\%$  of the events passing the selection cuts defined in Section \ref{sec:datareduction}. This value was corrected by a factor 0.8 determined using the MC simulations to take into account the different detection efficiencies of the 1.27 MeV photon in the case of the EC process and in the case of the transition with the positron. The measured value was consistent with the expected fraction of electron capture events (decay mode B). The cut $E_{\textrm{tot}}<80$ keV corresponding to the region containing 99$\%$ of the events in the EC peak at zero energy was used to define the photonless events.

To determine the signal inefficiency, mostly due to pileup, $\simeq 10^6$ events have been collected using a random trigger formed by delaying the fiber coincidence by 16 $\mu$s (half of the mean time interval between two events).  For the 80 keV threshold defined above, this gave an inefficiency of $(11.6 \pm0.5)\%$ that was consistent with the prediction of the simulation. This inefficiency was measured at the beginning of the data taking and, conservatively, was not corrected for the reduction of the source intensity during that period. 
   
For the FBGO, the broadening of the pedestal due to the contribution of energy deposited in the fiber had to be considered. The energy of the two 511 keV annihilation photons was localized in two opposite-located crystals of the barrel and then the FBGO pedestal was built. This resulted in a correction of 1$\%$ in the efficiency of zero signal detection. 

Finally, the signal efficiency was estimated to be $\epsilon \simeq (87.4 \pm0.5)\%$.

 The mean fraction of \ops\ in the data sample could be evaluated 
from the decay time curve by fitting the observed distribution to the 
function $A \cdot e^{(-t/\tau_{\ops})} + B$ ($B$ is the accidental background)
starting from the time $t=100$ ns when \ops\ was completely 
thermalized in the target \cite{thesis}. After taking into account the estimated difference of efficiency for 2 and 3 gamma detection and the pick-off effect (measured from the lifetime spectra with the same method as described in  Ref.\cite{rubbia}), the fraction of \ops\ in the data sample was $4.5\pm 0.2~\%$ \cite{thesis}. 
Since in the signal region no zero energy events were observed 
the upper limit for the branching ratio \cite{PDG} is: 

\begin{equation}\label{eq:BR_opsinv}
Br(o-Ps \rightarrow invisible) = 2.3/ ( N_{o-Ps} \cdot \epsilon) \leq 4.2\times10^{-7}~(90\%~\textrm{C.L.}) 
\end{equation}

where $N_{\ops}=(6.31\pm0.28) \times 10^6$ is the number of \ops\ in the selected sample.
Figure~\ref{EcalSum} shows the extrapolation into the region of the zero signal with an exponential.
The integral from 0 to 80 keV of the function obtained from the fit gives an evaluation of the background contribution in this region. The result is $N_{\textrm{bkg}}=0.34\pm 0.04$ expected events, where 
the error was evaluated from the uncertainty related to the 
extrapolation procedure itself.

This experiment can also be used to obtain upper limits on  $Br(\pps \to invisible)$ and on $Br(e^+e^-\to invisible)$.
For this purpose the different probabilities of positrons stopping in the fiber and in the aerogel were calculated with the help of simulations.
Several papers were reviewed in  Ref.\cite{mits} concluding that the probability to form positronium (75$\%$ of o-Ps and 25$\%$ of p-Ps) per positron stopping  is about 0.45 in the aerogel and between 0.2 and 0.4 in the fiber. In the aerogel a fraction of 0.05 to 0.1 of the p-Ps atoms experience pick-off annihilation in about 2 ns before escaping out of the silica grains in the pores. For o-Ps this probability ranges from 0.28 to 0.45. In a plastic scintillator (the fiber) the corresponding pick-off probabilities are 0.99 to 1 for \ops\ and 0.05 to 0.1 for p-Ps \cite{mits}. From the simulations, the fraction of positrons stopping in the fiber is 0.43 and in the aerogel is 0.25, therefore, the smallest possible fraction of p-Ps decays (including pick-off) is $5.5\%$. Thus, an upper bound for the invisible decay of p-Ps can be calculated:
\begin{equation}
Br(p-Ps\to invisible) = 2.3/(N_{p-Ps} \cdot \epsilon) \leq 4.3 \times 10^{-7}~(90\%~\textrm{C.L.})
\end{equation}
where $N_{p-Ps}=0.055\times 1.41\times 10^{8}\simeq 6.14\times 10^6$ was used and the positrons that do not stop in the fiber or in the aerogel are assumed to annihilate directly.\\
The number of $e^+e^-$ decays can be calculated subtracting from the total number of events the p-Ps and o-Ps decays, thus, one obtains $N_{e^+e^-}\simeq 1.29\times 10^8$ and an upper limit for the branching ratio of: 
\begin{equation}
Br(e^+e^-\to invisible) = 2.3/( N_{e^+e^-} \cdot \epsilon) \leq 2.1 \times 10^{-8}~(90\%~\textrm{C.L.})
\end{equation}

\section{Interpretation}\label{sec:interpretation}  

No event consistent with an invisible decay was found in the large sample of events.

\begin{figure}[!h]
 \begin{center}
\includegraphics[width=.45\textwidth,height=.4\textwidth]{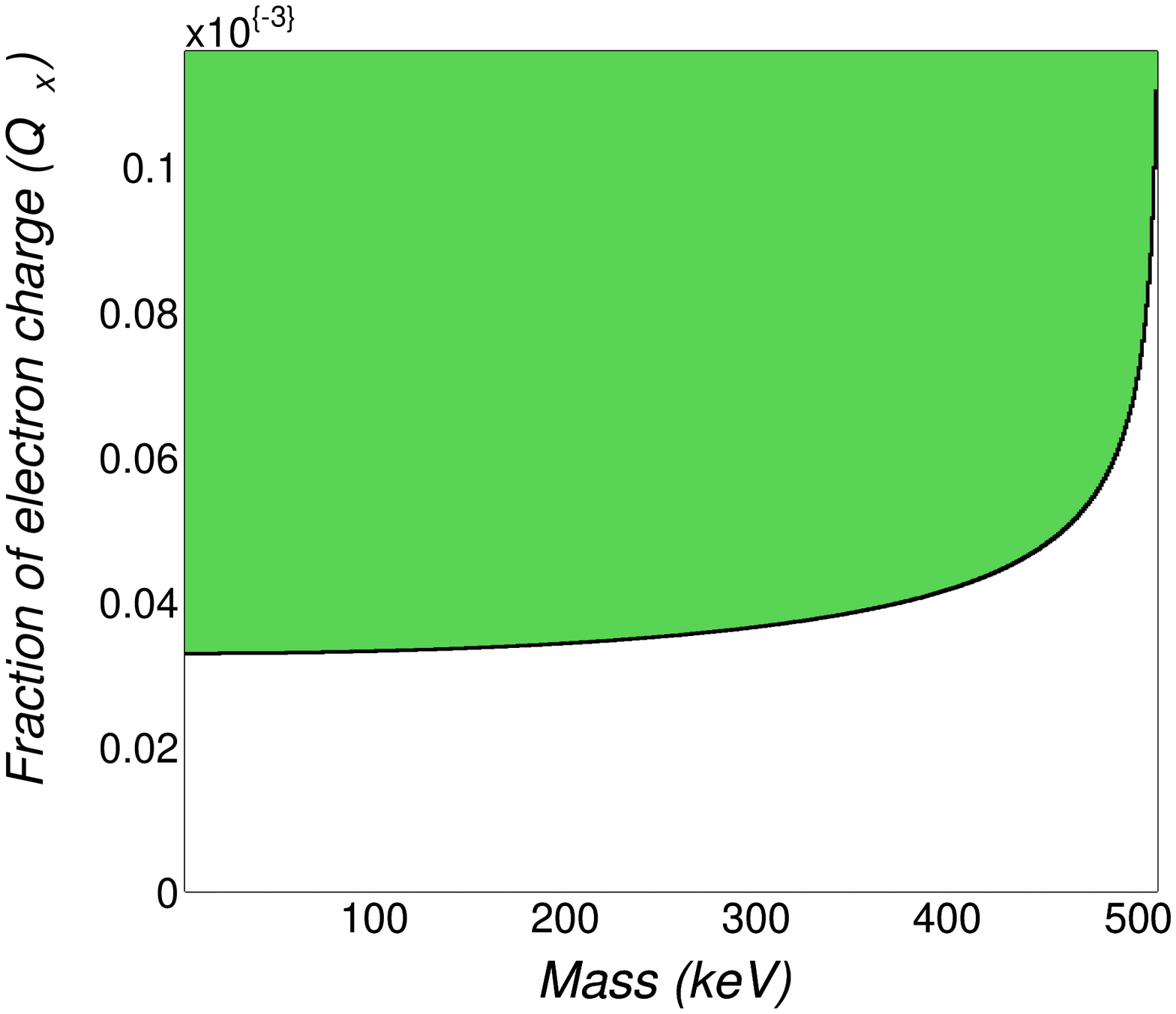}
\includegraphics[width=.45\textwidth]{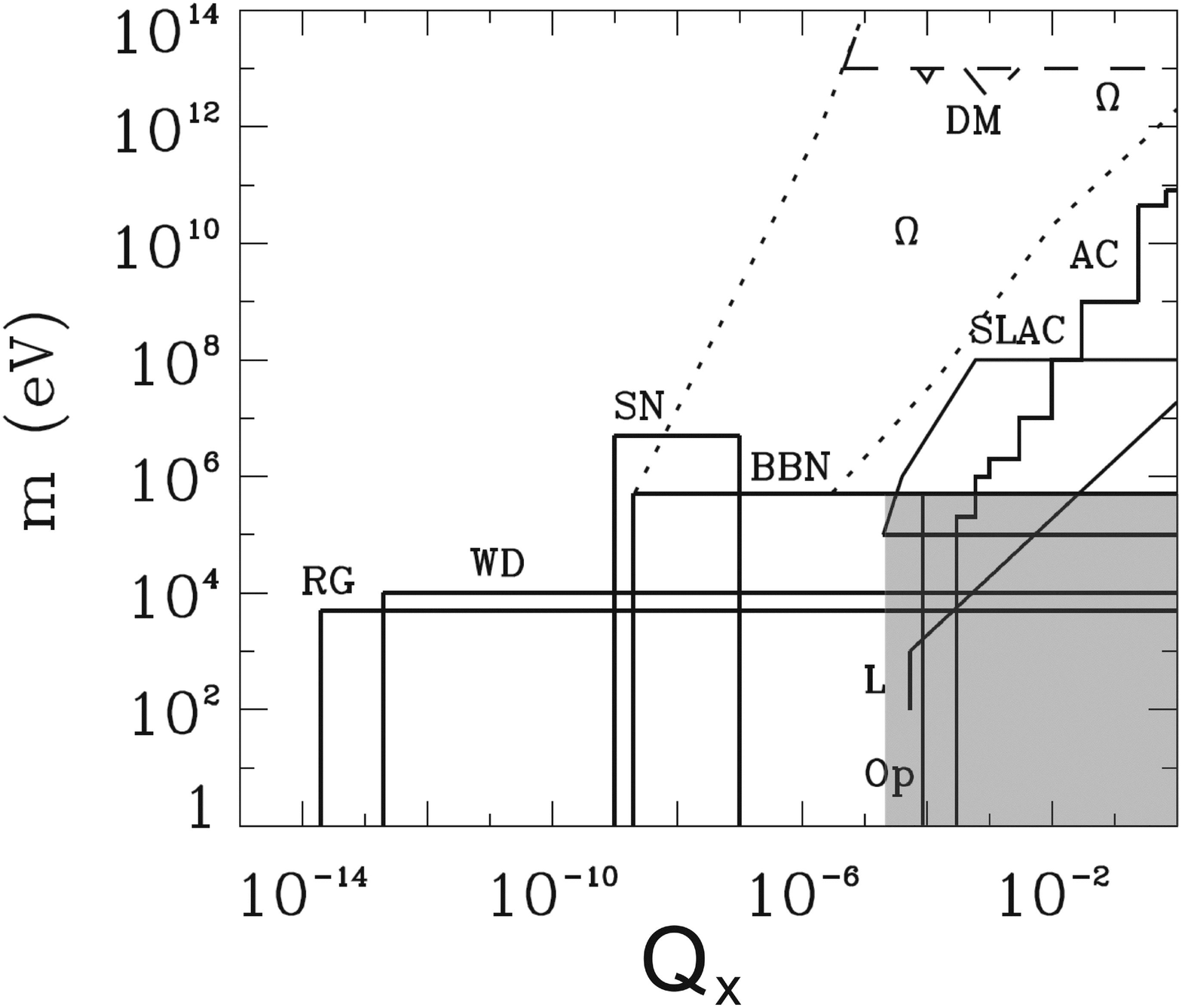}
\caption{\em a) Mass--charge parameter space for the \ops\ decay into milli-charged particles, excluded with this experiment, b) Comparison of our results (the dashed region on the plot) with other experimental (SLAC \cite{prinz} and previous \invdecay\ \cite{mits}) and astrophysical bounds (the plot was taken from \cite{david}).}
\label{milli}
 \end{center}
 \end{figure}

Using Eq. (32) of Ref.\cite{extradim}, the bound for particles with a fraction  $Q_x$ of the electron charge can be plotted as a function of their mass $m_X$ for $m_X<m_e$, as shown in Fig.~\ref{milli}a. Thus, the region of the charge-mass parameter space, which was not excluded directly by the SLAC results \cite{prinz} and the previous search for \invdecay\ \cite{mits},  is covered by this experiment (see Fig.~\ref{milli}b).

The strength of the photon mirror-photon mixing $\epsilon$ can be extracted from the limit on the \binvdecay\ with \cite{gninenko}:
\begin{equation}\label{eq:mixing_mirror_cavity}
\epsilon= \frac{1}{2\pi f}\sqrt{\frac{Br(\invdecay)\Gamma_\mathrm{SM}\Gamma_\mathrm{coll}}{2(1-Br(\invdecay))}}\end{equation}
by substituting a conservative value of $\Gamma_\mathrm{coll}=5\times 10^{4}~\mathrm{s}^{-1}$ for the collision rate of the \ops\ against the walls of the aerogel pores \cite{foot}. $\Gamma_\mathrm{SM}$ is the decay rate of \ops\ in vacuum and  $f = 8.7\times 10^4$ MHz is the contribution to the
ortho-para splitting from the one-photon annihilation diagram
involving \ops\ \cite{glashow}. Using our result one can estimate the mixing strength to be $\epsilon \leq 1.55\times 10^{-7}~(90\%~\textrm{C.L.})$. This is close to the BBN limit of $\epsilon < 3\times 10^{-8}$ \cite{cg} but does not cover all the region of interest suggested by the DAMA and CRESST results \cite{foot1} and motivated by GUT predictions \cite{berezhiani} and by string theory \cite{abel} ($\epsilon > 10^{-9}$).

%An experiment in vacuum \cite{wsinvisible} could exclude this possibility completely since the collision rate will be 4-5 orders of magnitude smaller.

\section{Conclusion}\label{sec:conclusion}

In this paper the results of a new search for an invisible decay of \ops\ were reported.  Since no event was found in the energy window [0,80] keV,  an upper limit for the branching ratio was set:
\begin{equation}
 \binvdecay \leq 4.2\times 10^{-7}~(90\%~\textrm{C.L.}) 
\end{equation}
improving the best existing bound \cite{mits} by a factor 7.  

Analyzed in the context of theoretical models the negative result provides an upper limit on the photon mirror-photon mixing strength  $\epsilon \leq 1.55\times 10^{-7}~(90\%~\textrm{C.L.})$ and rules out particles with a fraction  $Q_x \leq 3.4 \times 10^{-5}$ (for $m_X \leq m_e$) of the electron charge (milli-charged particles). Furthermore, an upper limit on the branching ratios for the process $Br(p-Ps\to invisible)\leq 4.3 \times 10^{-7}~(90\%~\textrm{C.L.})$ and $Br(e^+e^-\to invisible)\leq 2.1 \times 10^{-8}~(90\%~\textrm{C.L.})$ could be set.
 
{ \bf Acknowledgements}\\

We thank the Paul Scherrer Instititute (Villigen, Switzerland) for lending us the BGO crystals.  We gratefully acknowledge the help of
 B. Eichler and J. Neuhausen for the $^{22}$Na source preparation. We wish to thank  Z. Berezhiani, R. Eichler,  S. Karshenboim, N.V. Krasnikov, V. Matveev and V.A. Rubakov for  useful  discussions. We are grateful to N.A. Golubev, L. Knecht, G. Natterer, J. P. Peigneux and M. Sauter for their essential help. We wish to thank M. Haguenauer for providing us with samples of scintillating fibers.
This work was supported by the Swiss National Science Foundation and by the French Ministery of Foreign Affairs through an ECONET program. We thank the CERN for its hospitality.

\end{document}